# Driving force induced transition in thermal behavior of grain boundary migration in Ni

Xinyuan Song and Chuang Deng*

Department of Mechanical Engineering, University of Manitoba, Winnipeg, MB R3T 2N2, Canada

* Corresponding author: Chuang.Deng@umanitoba.ca

**Abstract**

Grain boundary (GB) migration exhibits intriguing anti-thermal behavior (or non-Arrhenius behavior), with the temperature and driving force playing crucial roles. Through atomistic simulations on nickel bicrystals, we investigate the change in GB mobility with variations in both temperature and driving force. Our results reveal that the GB mobility initially increases with temperature and subsequently decreases after reaching the transition temperature ($T_{trans}$), and, notably, $T_{trans}$ exhibits a linear relationship with the activation energy ($Q$) associated with GB migration. By modulating the driving force, we found that the driving force could effectively lower $Q$, resulting in the shift of $T_{trans}$ towards lower temperatures. Additionally, higher driving forces were found to activate more migration modes at lower temperatures, potentially leading to a transition in the thermal behavior of GB migration. Our work supports the existing theoretical models for GB migration based on both classical thermal activation and disconnection nucleation. Furthermore, we refined the existing model by incorporating the influence of the driving force. The modified model can not only describe the effect of driving force on the thermal behavior of GB migration but also accounts for the observed "anti-driving force" phenomenon in GB migration. Our research has the potential to offer valuable insights for investigating realistic GB migration under more intricate constraints and environments.

## 1. Introduction

Grain boundary (GB) mobility is a fundamental dynamic property that characterizes the rate at which a GB moves in response to an external driving force, which is an important parameter that influences the microstructural evolution of polycrystalline materials. Despite the apparent physical meaning of GB mobility and extensive investigations conducted through experiments [1–4] and atomistic simulations [5–14], accurately predicting GB migration behavior remains challenging due to the complexity arising from the five-parameter space associated with GBs. While numerous theoretical models, such as those based on the conventional concepts of the structural unit[15,16] and the more recent ones based on machine learning and the various types of local atomic descriptors[17,18], have been developed to successfully predict other fundamental properties of GBs, such as GB energy[18] and the energy spectrum for solute segregation[17], no such model can accurately forecast the migration behavior of GBs, even for GB structures of high symmetry. Therefore, a profound understanding of GB migration behavior and the corresponding predictive theory are highly desired in the community and hold significant importance for modern manufacturing industries[19].

The migration of GBs has traditionally been considered thermally activated, following the empirical Arrhenius relation [4,20]: $M = M_0 \exp(-Q/k_B T)$, where $M_0$ is a constant prefactor, $Q$ is the activation energy, $k_B$ is the Boltzmann constant, and $T$ is the temperature. However, in recent years, the phenomenon of anti-thermal behavior (or non-Arrhenius behavior) in GBs has been widely reported, which refers to the decrease in GB mobility with increasing temperature. For instance, atomistic studies [10–12] have indicated the prevalence of anti-thermal behavior in specific GBs, such as Σ3, Σ7, and Σ9 GBs in Ni. Experimental observations, as discussed in [4], have also reported faster GB migration at cryogenic temperatures [21] and slower GB migration at higher temperatures [22,23].

The mechanism underlying the anti-thermal behavior of GB migration remains a subject of ongoing debate. One hypothesis, put forth by Priedeman et al. [10], suggests that the slowdown in GB migration can be attributed to phonon drag, which is the mechanism that hinders the motion of dislocation. However, a recent study conducted by Homer et al. [24] presents a different perspective on the anti-thermal behavior of GB migration. They reintroduce the classical thermal-activation model originally proposed by Gottstein and Shvindlerman [25] and argue that the observed deceleration of GB migration at lower temperatures is a thermal activation phenomenon rather than an external influence. Additionally, Chen et al. [7] propose a theoretical model based on disconnection theory, providing further support for the notion that the anti-thermal behavior is an inherent characteristic of GBs. Several other hypotheses have also been put forward, such as structural phase transitions [26,27], roughening transitions [8,28], and topological phase transitions [29]. However, further investigations are needed to substantiate these hypotheses.

The classical kinetic equation for GB migration is given by:

$$v = MP|_{P \to 0} \tag{1}$$

where $v$ is the velocity of the GB motion and $P$ is the driving force, and $M$ is the GB mobility. According to this equation, the GB mobility should be independent of the driving force when the driving force approaches zero [30]. However, in previous studies using molecular dynamics (MD) simulations, relatively large driving forces were applied to observe GB migration within nanosecond timescale. This approach have a great effect on the measured mobility [11,12] and thermal behavior of GBs [11]. For instance, as reported in [11], when the applied driving force for the Ni Σ7 (5 4 1)/(5 4 $\bar{1}$) GB (P26 in the Olmsted database[31]) changes from 74 to 372 MPa, the GB migration behavior transitions from thermally activated to anti-thermal. With the same change in driving force, the Σ7 (8 5 1)/(7 5 4) GB (P207 in the Olmsted database[31]) exhibits a transition from anti-thermal to thermally activated migration behavior. This dependency of thermal behavior on the driving force complicates the study of thermal behavior of GB and direct comparisons between experimental observations (typically performed with driving forces in the range of $10^2$ - $10^6$ Pa[32]) and atomistic simulations (with driving forces typically larger than $10^7$ Pa [11]). To date, no systematic research has been conducted to investigate the effect of the driving force on the thermal behavior of GB migration, further emphasizing the need for comprehensive studies in this area.

Trautt et al. [9] have proposed a method to extract GB mobility at the zero-driving force limit from the random walk of the GB. Deng and Schuh [33] further enhanced the accuracy of this method, enabling it to capture the subtle movements of GBs more effectively. According to this approach, the mobility of a flat and fully periodic GB can be computed using the Einstein relation [9]:

$$D = 2Mk_BT/A \tag{2}$$

where *A* represents the GB area, *D* is the GB diffusion coefficient. It is important to note that *D*, in this context, is based on the mean square displacement (MSD) of the average normal migration of the entire GB plane. This is distinct from the conventional GB diffusivity, which tracks the MSD of individual GB atoms, although these two properties may be fundamentally correlated. At low temperatures, GBs are expected to exhibit neglectable movement due to the limited thermal fluctuations. As a result, the random walk method has primarily been utilized to compute GB mobility in the high-temperature regime (typically above 500K) in previous studies [9,34]. However, in a recent study by Homer et al. [24], it was demonstrated that the anti-thermal behavior of GB migration is likely associated with small activation energies for GB migration, which implies that random walk of these GBs can be obvious at low temperatures.

In this paper, we aim to investigate the thermal behaviors of GBs with different activation energies using the interface random walk method. This approach will help determine whether the anti-thermal GB migration is an intrinsic property of GBs or a thermal phenomenon induced by external driving forces. Furthermore, by using the mobility and thermal behavior of GBs obtained through the random walk method as a reference, the study systematically explores the influence of external driving forces on the activation barriers for GB migration and the thermal behavior of GBs. The findings of this investigation are subsequently compared with existing theoretical models, and the effects of driving forces are discussed. The aim of this study is to shed light on the physical mechanisms underlying the thermal behavior of GB migration and illustrate how external driving forces can play a role.

## 2. Method

*2.1 GB dataset*

In order to investigate the relationship between activation energy and the thermal behavior of GB migration, we have selected 9 coincidence site lattice (CSL) Ni GB models from the Olmsted database [31]. These GB models are known to potentially exhibit small activation energies for migration based on previous studies [11,35]. Two methods are employed to compute the activation energy of the GBs: The first method involved fitting the $\ln M$-$1/k_B T$ line to calculate the apparent activation energy, denoted as $Q$. This method will be discussed in detail in Section 4.2. The second method used nudged elastic band (NEB) methods [36–38] to compute the energy barrier during GB migration, denoted as $E$. See Supplemental Material for detailed procedure for computing the energy barriers [36–41]. This method has been widely used in previous research on GB migration [7,42,43]. Fig. 1 illustrates a typical GB model, and Table 1 provides detailed information about each GB. To study the "intrinsic" thermal behavior of GB migration and the influence of driving force, the boundary condition in the $x$ direction is set as shrink-wrapped (free surfaces) to eliminate constraints and other factors that could affect GB migration, except for temperature and driving force. The boundary conditions in the $y$ and $z$ directions (GB plane directions) are set as periodic. All simulations are performed using the Large-scale Atomic/Molecular Massively Parallel Simulator (LAMMPS) package [44] with an embedded atom method (EAM) potential [45]. The simulations are conducted within the temperature range of 10K to 800K, and in some cases, up to 1000K.

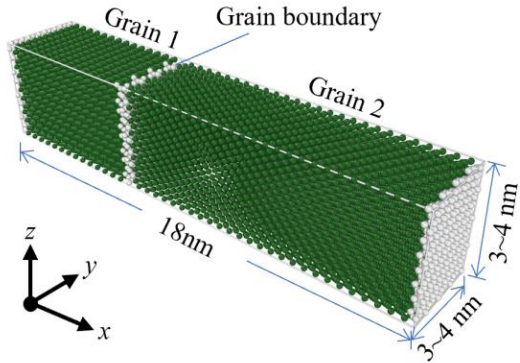

Figure 1 A representative atomistic model of the simulated Ni GBs. Atoms located at perfect FCC sites and non-FCC sites are color-coded in green and white, respectively.

Table 1 The detailed information on the GBs investigated in this study, specifically their energy barriers for GB migration $E$ and apparent activation energy $Q$. The $E$ are presented in ascending order from low to high values. Some GBs show two energy barriers during the migration ($E$) which will be discussed in Section 4.2.

| ID in the Olmsted database[31] | Sigma | GB plane | Energy barrier for GB migration $E$ (eV/nm$^2$×10$^{-3}$) | Apparent activation energy $Q$ (eV) |
| --- | --- | --- | --- | --- |
| 207 | 7 | (8 5 1)/(7 5 4) | 0.61 | 0.0024 |
| 256 | 87 | (11 7 2)/(11 7 2) | 2.36 | 0.0054 |
| 87 | 9 | (7 5 4)/(7 5 $\bar{4}$) | 2.84 | 0.0076 |
| 57 | 35 | (4 2 0)/(4 2 0) | 3.37 | 0.0078 |
| 167 | 35 | (10 6 2)/(10 6 2) | 4.73 | 0.027 |
| 257 | 29 | (11 7 2)/(11 7 $\bar{2}$) | 5.42 | 0.046 |
| 25 | 21 | (5 4 1)/(5 4 1) | 7.97 | 0.072 |
| 30 | 3 | (3 2 1)/(3 2 $\bar{1}$) | 0.29/5.08 | Almost 0 |
| 14 | 15 | (2 1 1)/(2 1 1) | 1.89/6.67 | 0.018 |

*2.2 Random walk simulation*

Before tracking the GB migration, the models were expanded at different temperatures based on the corresponding thermal expansion coefficient. Subsequently, equilibration was performed under the isothermal-isobaric ensemble (NPT) for 10 ps, followed by a short annealing period of 5 ps under the microcanonical ensemble (NVE) using the Berendsen thermostat. The Berendsen thermostat was then removed, and the GBs were left to fluctuate randomly for 5 ns solely under the thermal effects in the NVE ensemble. The variation in the order parameter, as described in reference [46], was employed to track the mean position across the GB plane within the system. Additionally, to determine the shear coupling factor during the thermal fluctuations of each GB, the relative shear displacement in each model was recorded, by tracking the center of mass of a thin slab, approximately 1 nm thick, at both ends along the *x* direction. To ensure the reliability of the results, each simulation was repeated 20 times with different random seeds for the initial velocity distribution. The duration of each simulation was 5 ns, and the resulting data sets were split into 10 groups of 500 ps, resulting in a total of 200 sets of independent simulation results for each condition. Fig. 2 illustrates an example of GB migration across 200 simulations. The initial GB position was defined as 0 for all simulations. It was observed that even at a low temperature of 200K, the GB with a Σ15 (2 1 1) structure exhibited detectable migration solely due to thermal effects. The GB migration was found to be entirely random, with the average position (indicated by the thick black line) remaining in the middle.

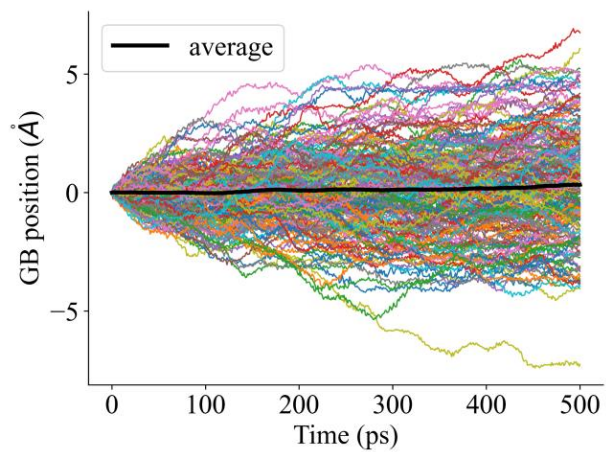

Figure 2 The migration of Σ15 (2 1 1) GB under the effect of thermal fluctuation at 200K.

For each simulation, the cumulative distribution function $f(x)$ [33] at a given time $t$ can be described by the following equation:

$$f(x) = \frac{1}{2}\left[1 + \text{erf}\left(\frac{x - \mu}{\sigma\sqrt{2}}\right)\right] \qquad (3)$$

where $f(x)$ is the probability that the displacement of the mean position across the GB plane, denoted as $d(t)$, falls within the range $(\infty, x]$, and erf is the error function. The fitted parameters $\mu$ corresponds to the average displacement $\langle \bar{d}(t) \rangle$, and $\sigma^2$ corresponds to the mean square displacement $\langle \bar{d}^2(t) \rangle = Dt$. Therefore, $D$ can be calculated by $d\sigma^2/dt$, and the GB mobility $M$ can be determined using Eq. 2.

*2.3 External driving force driven GB migration*

The effect of the driving force on GB migration was investigated using the energy-conserving orientational (ECO) synthetic driving force [46,47]. As a point of comparison, the influence of shear stress was also examined by applying opposing shear forces on the two thin slabs at both ends along the *x* direction. Each GB was subjected to this driving force until it reached one end of the model, with a maximum simulation time of 5 ns. To ensure statistical significance, each simulation was repeated 20 times with different random seeds for initial velocity distribution.

The velocity of the GB in each simulation was determined by linearly fitting the displacement vs. time curve through the least square error method[48] and calculating the slope of the fitted line. The final GB velocity was obtained by averaging the velocities from the 20 simulations. Based on the calculated velocity, the GB mobility was then determined using Eq. 1.

**3. Results**

*3.1 Temperature-dependent GB mobility: random walk simulations*

In the random walk simulations, all 9 GBs listed in Table 1 exhibited anti-thermal behavior in terms of GB mobility. The paper specifically focuses on three representative GBs, namely Σ15 (2 1 1), Σ9 (7 5 4)/(7 5 $\bar{4}$), and Σ21 (5 4 1) GBs, and their respective *M* vs. *T* curves are shown in Figs. 3a-c (See Supplemental Materials for the *M* vs. *T* curves of the remaining GBs). The results indicate that at low temperatures, the GB mobility displayed a thermally activated behavior, where higher temperatures led to increased mobility. However, as the temperature further increased, a transition occurred, and then the GB mobility exhibited anti-thermal behavior, as shown in Fig. 3d. This transition point, represented as $T_{trans}$, marks the temperature at which the thermal behavior of GB migration undergoes a transition.

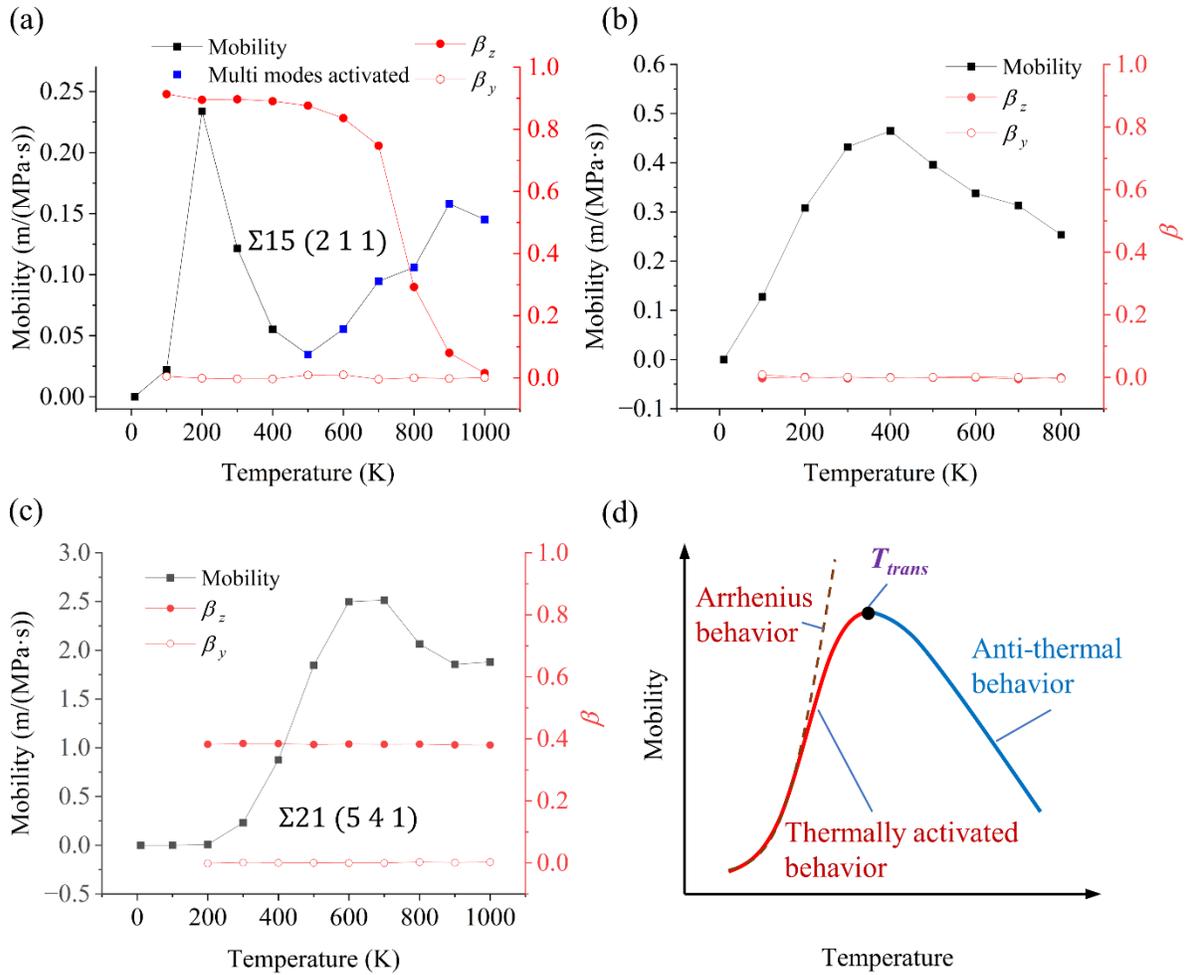

Figure 3 The plots of GB mobility and shear coupling factor vs. temperature determined from random walk simulation for (a) Σ15 (2 1 1) (the blue dots indicate a change in disconnection modes at specific temperatures), (b) Σ9 (7 5 4)/(7 5 $\bar{4}$) and (c) Σ21 (5 4 1) GBs. (d) Illustration of the thermally activated behavior (i.e., mobility increases with temperature), the anti-thermal behavior (i.e., mobility decreases with temperature), and the transition temperature $T_{trans}$.

According to the unified GB kinetics model proposed by Han et al. [30], GBs can migrate through different modes, each characterized by a specific shear coupling factor $\beta$. The dichromatic pattern analysis, presented in the Supplemental Materials, reveals that the migration modes of Σ15 (2 1 1) and Σ21 (5 4 1) GBs exhibit different $\beta$. Consequently, monitoring the changes in $\beta$ allows us to track the activation status of migration modes for Σ15 (2 1 1) and Σ21 (5 4 1) GBs during the migration process, as demonstrated in Figs. 3a and c. On the other hand, for Σ9 (7 5 4)/(7 5 $\bar{4}$) GB, its migration modes either do not possess a shear coupling effect or consist of two similar modes with opposite $\beta$, indicating a final $\beta$ of 0. which is consistent with the observation shown in Figs. 3b.

For both Σ15 (2 1 1) and Σ21 (5 4 1) GBs, it was observed that only a single mode (indicated by the constant $\beta$) was detected around the $T_{trans}$. This indicates that the transition in thermal behavior is not caused by a change in the migration mode. Specifically, for the Σ15 (2 1 1) GB, the shear coupling factor begins to decrease at 500 K, suggesting the activation of multi-modes at this temperature. This may be the underlying mechanism that leads to an increase in GB mobility after 500 K (as indicated by the blue data points in Fig. 3a) and another transition in thermal behavior, specifically from anti-thermal to thermally activated behavior.

*3.2 The effect of driving force*

The effect of the synthetic driving force on GB mobility is illustrated in Fig. 4. The large error bars observed at small driving forces indicate the stochastic nature of GB migration caused by random thermal fluctuations, as shown in Fig. 2. At low driving forces, such as 1.47 MPa, the GB mobility calculated using Eq. 1 closely matches that determined by the random walk method, consistent with previous findings [34]. As the driving force increases, the transition temperature $T_{trans}$ for the thermal behavior of GB mobility shifts towards lower values. Notably, for Σ15 (2 1 1) and Σ9 (7 5 4)/(7 5 $\bar{4}$) GBs, the first peak in mobility completely disappears at driving forces of 58.8 MPa and 29.4 MPa, respectively. Particularly for the Σ9 (7 5 4)/(7 5 $\bar{4}$) GB, a monotonic anti-thermal trend is observed when it is driven by this force up to 800 K. When the driving force reaches an extremely large value of 440.97 MPa, the mobility of Σ9 (7 5 4)/(7 5 $\bar{4}$) and Σ21 (5 4 1) GBs becomes essentially independent of temperature, in agreement with the ballistic transition proposed by Deng and Schuh [32]. However, for Σ15 (2 1 1) GB, the mobility reverts back to a thermally activated behavior. This behavior requires further investigation and will be discussed in more detail below. Supplemental Figure S4 provides the *M* vs. *T* curves for driving forces ranging from 58.8 to 440.97 MPa, as well as the *M* vs. *P* curves at different temperatures.

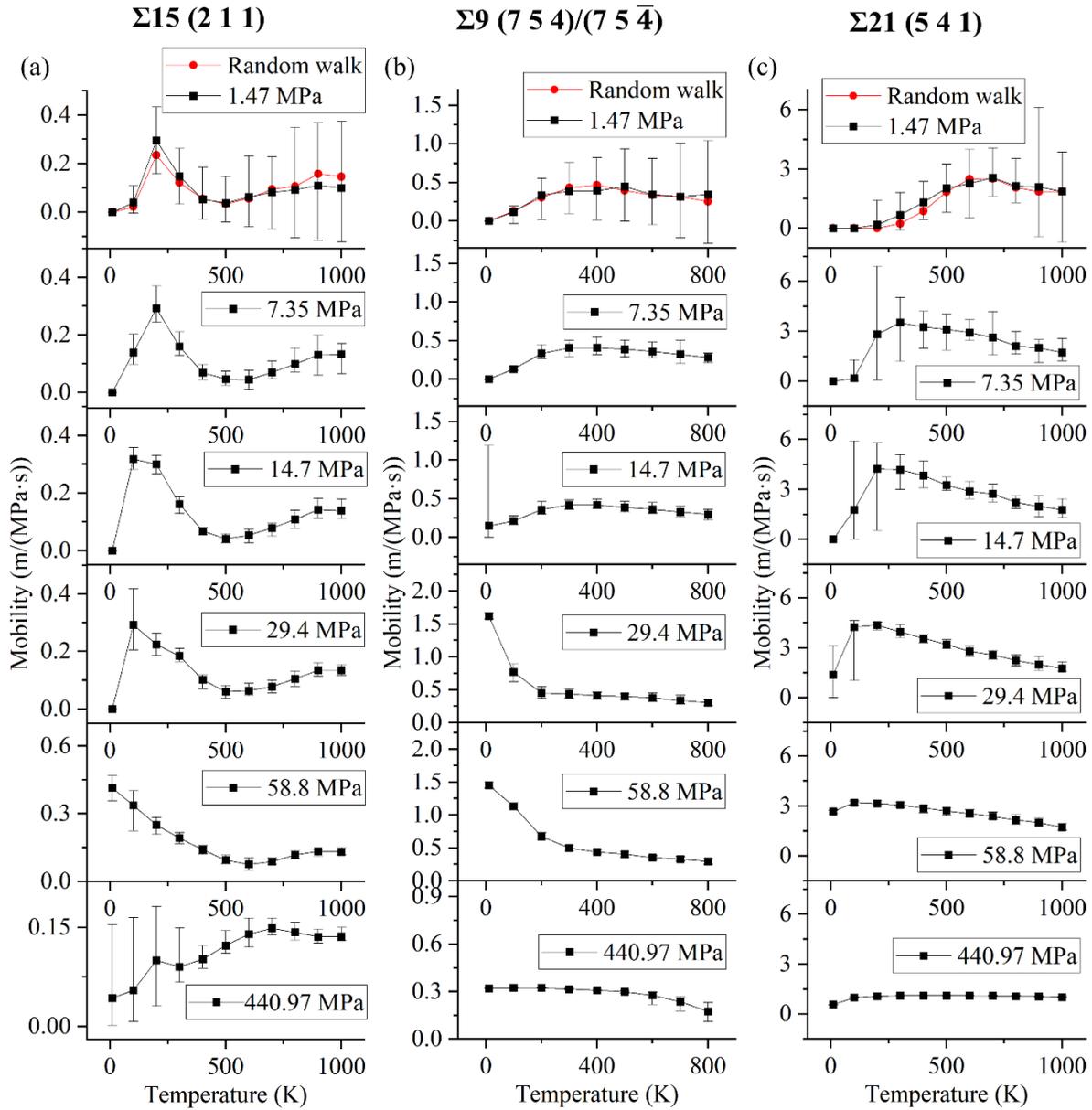

Figure 4 The effect of synthetic driving force on GB mobility for (a) Σ15 (2 1 1), (b) Σ9 (7 5 4)/(7 5 $\bar{4}$), and (c) Σ21 (5 4 1) GBs in Ni.

To investigate the underlying mechanisms behind the transition in thermal behavior of GB migration, the $\beta$ for the Σ15 (2 1 1) GB under different driving forces and temperatures was analyzed and presented in Fig. 5. The heatmap reveals that both a larger driving force and higher temperature can activate more migration modes during GB migration. Interestingly, when the synthetic driving force reaches 440.97 MPa, multiple modes can be activated even at very low temperatures, as low as 10 K. This observation may explain why the GB mobility of Σ15 (2 1 1) changes back to a thermally activated behavior (Fig. 4a) under the influence of a driving force of 440.97 MPa.

It is important to note that the $T_{trans}$ in thermal behavior (Fig. 4a), occurring around 200 K and below for the cases when driving force ≤ 58.8 MPa, do not correspond to the transition in migration modes shown in Fig. 5 (indicated by red line). A similar phenomenon was also observed for the Σ21 (5 4 1) GB (Fig. 3c), in which there is no change in $\beta$ across all temperatures (10 K to 1000 K), yet a peak in the GB mobility emerges around 700 K. Therefore, the transition from thermally activated to anti-thermal behavior in GB migration, as well as its dependence on driving force, cannot be solely attributed to the change in migration modes that are activated during the migration process.

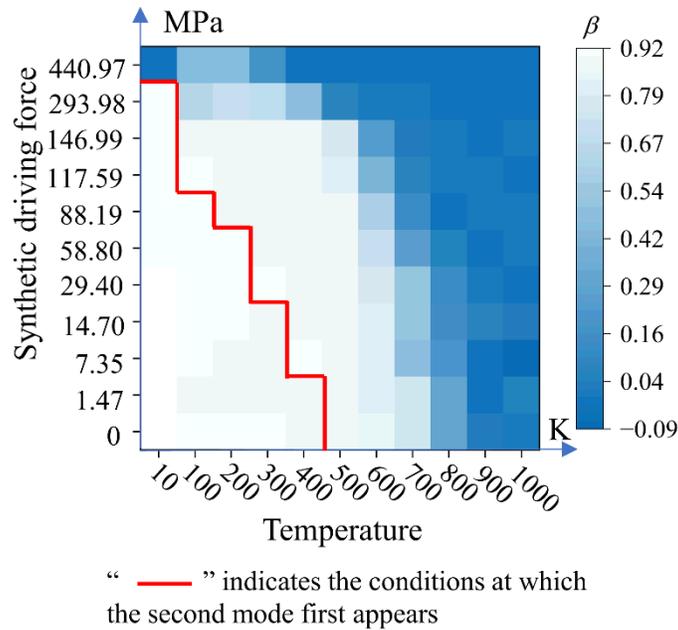

" ——— " indicates the conditions at which the second mode first appears

Figure 5 Heatmap of the shear coupling factor in the Σ15 (2 1 1) Ni GB under varying driving forces and temperatures

In addition, another interesting observation is the presence of an "anti-thermal" trend in GB migration, or more accurately, an "anti-driving force" trend, with increasing external driving forces. Fig. 6 illustrates this behavior, where at a low temperature of 100 K, the mobility M of Σ15 (2 1 1) GB initially increases and then decreases as the driving force increases. This trend is reminiscent of the anti-thermal phenomenon observed in GB mobility when the temperature changes under low driving force conditions (Fig. 4). In contrast, at a high temperature of 800 K, the GB mobilities remain nearly constant and do not exhibit a significant dependence on the driving force. A similar driving force-dependent "anti-thermal" trend in GB mobility has been reported in previous studies, such as those by Deng and Schuh [32] and Race et al. [49]. This intriguing phenomenon will be further discussed in the subsequent sections.

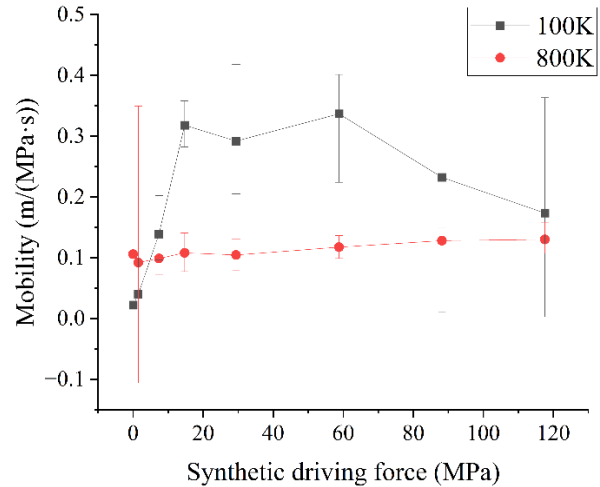

Figure 6 Mobility vs. driving force curves at 100K and 800K for Σ15 (2 1 1) GB

## 4. Discussion

### 4.1. Theoretical models for anti-thermal GB migration

The classical thermal-activation model, originally proposed by Gottstein and Shvindlerman [25] and recently reintroduced by Homer et al. [24], offers an explanation for the anti-thermal behavior observed in GB migration. In this model, atoms randomly jump forth and back across the GB based on transition state theory, and a collective atom jump leads to GB migration. Fig. 7a illustrates the model, where $Q$ represents the activation energy for the atomic configuration transformation from $S_1$ to $S_2$, and $\Psi$ represents the energy introduced by the external driving force. The velocity of the GB can be calculated using the equation:

$$v = Nb\left[\omega^+ \exp(-\frac{Q}{k_BT}) - \omega^- \exp(-\frac{Q+\Psi}{k_BT})\right]$$

$$= Nb\omega \exp\left(-\frac{Q}{k_BT}\right)\left[1 - \exp\left(-\frac{\Psi}{k_BT}\right)\right] \quad (4)$$

where, $N$ is the number of atoms, $b$ is the distance of the atom jump, $\omega$ is the attempt frequency, + and - denote the forth and back directions of the GB migration. As the external driving force approaches 0, the terms in the square bracket in in Eq. 4, i.e. $1-\exp(-\Psi/k_BT)$ can be approximated as $\Psi/k_BT$. Since $\Psi$ is the energy drop introduced by the external driving force and is proportional to the driving force $P$ ($\Psi = CP$, where $C$ is a constant), the GB mobility in the zero-driving force limit can be derived from Eq. 4 as:

$$M = \frac{v}{P} = \frac{Nb\omega C}{k_BT} \exp\left(-\frac{Q}{k_BT}\right) \quad (5)$$

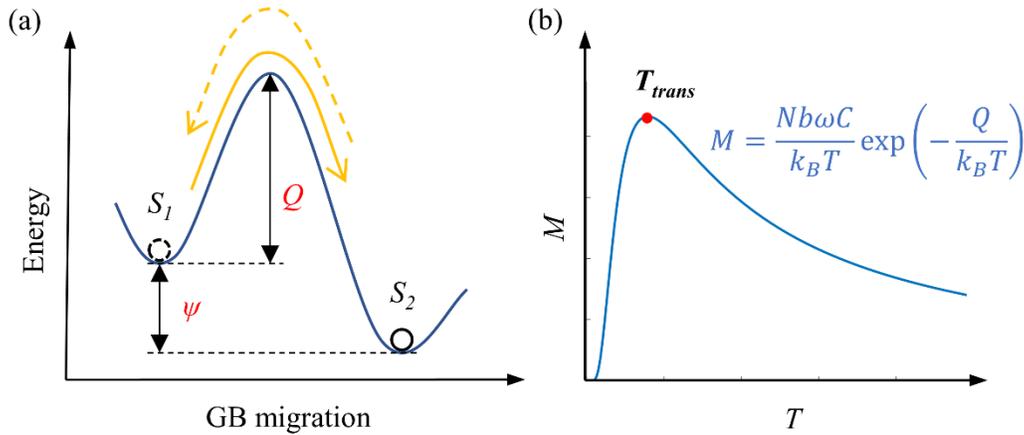

Figure 7 (a) Illustration of the classical thermal-activation model [24] and (b) Plot of the mobility vs. temperature curve based on Eq. 5

As shown in Fig. 7b, Eq. 5 provides a good description of the anti-thermal behavior of GB migration. The GB mobility $M$ initially increases with temperature $T$ and after reaching its maximum value at $T_{trans}$, $M$ starts to decrease as $T$ further increases. This trend is consistent with the observations depicted in Figs. (3) and (4).

Chen et al. [7] also proposed an equation to describe the anti-thermal GB migration based on the disconnection theory, taking into account the activation and migration of disconnections. The equation is given as:

$$M = \frac{2\omega\delta w}{k_B T} \sum_m \frac{h_m^2 \exp\left(-\frac{E_m^* + E_m^c}{k_B T/w}\right)}{1 + \frac{2}{e}\exp\left(\frac{E_m^e - 2E_m^c}{k_B T/w}\right)} \tag{6}$$

where $w$ is the thickness of a bicrystal, $\delta$ is the size of a CSL cell, $e = \exp(1)$ is Euler's number, subscript $m$ represents different disconnection modes, $h$ is the height of the disconnection, $E^C$ is the formation energy for a pair of disconnection, $E^*$ is the energy barrier for disconnection migration along the GB, and $E^e$ attributes to the long-range elastic interactions between disconnections. Due to the small magnitude of $E^e$ compared to $E^C$, the denominator of the second fraction in Eq. 6, i.e. $1+2/e\cdot\exp[E^e-2E^C/(k_BT)w]$, can be approximated as 1. Therefore, when only one disconnection mode is activated, Eq. 6 can be simplified as:

$$M = \frac{2\omega\delta w h^2}{k_B T} \exp\left(-\frac{E^* + E^C}{k_B T/w}\right) \tag{7}$$

Indeed, both Eq. 7 and Eq. 5 share a similar form, with the activation energy $Q$ in Eq. 5 corresponding to the combined energy barrier for disconnection nucleation $E^C$ and migration $E^*$ in Eq. 7.

Both theories indicate that the anti-thermal behavior of GB mobility is a result of the first term $1/k_B T$, which is arises from the inequality of the activation energy for GB to move forth ($Q$) and back ($Q+\Psi$) as illustrated in Fig. 7a. This inequality of the activation energy is inherent to GB migration and contributes to the anti-thermal behavior observed. This aligns with our observation that the transition in thermal behavior of GB mobility is not solely attributed to changes in the migration mode, as supported by the analysis of the shear coupling factor (Figs. 3 and 5).

*4.2. The activation energy for GB migration*

In Eq. 5, the activation energy $Q$ for GB migration plays a crucial role in deciding the thermal behavior of the GB migration. By setting the derivative of Eq. 5 with respect to $T$ equal to zero, i.e., $M'=0$, we can get

$$Q = k_B T_{trans} \tag{8}$$

Eq. 8 suggests a linear relationship between $Q$ and $T_{trans}$ by theory, which means that if we know the activation energy $Q$, the thermal behavior of the GB is predictable. But the question falls on how to determine the activation energy of the GB?

According to Chen et al. [7], the activation energy $Q$ in Eq. 5 is linked to the energy required for disconnection nucleation and migration (as shown in Eq. 7); the latter can be determined by performing NEB analysis [7,42,43]. Therefore, the energy barriers of 9 GBs listed in Table 1 are calculated through NEB method, and the maximum height of the energy barrier $E$ was extracted, as shown in Fig. 8a. It is important to note that the $E$ is reported as normalized values by dividing each energy barrier by the respective GB area to eliminate the possible size effect of NEB simulation.

Alternatively, by examining Eq. 5, one can deduce that

$$\ln M = -\frac{Q}{k_B T}\left(1 + \frac{k_B T \ln T}{Q}\right) + A \tag{9}$$

where $A$ is a constant. When $k_B T \ln T \ll Q$, i.e. $T \ll T_{trans}$, the terms in the bracket becomes 1, and the GB mobility $M$ exhibits Arrhenius behavior, as shown in Fig. 3d, and therefore, in this temperature range the apparent activation energy $Q$ can be calculated by fitting the slope of $\ln M$-$1/k_B T$ line. The calculated $E$ and $Q$ are listed in Table 1. Fig. 8b shows that, even though with different values, $E$ and $Q$ exhibit a linear relationship. Besides, both $E$ and $Q$ exhibit a linear relationship with $T_{trans}$ as predicted by Eq. 8, as shown in Figs. 8c, d. It is worth noting that the slope of fitted $Q$-$T$ line is $7.4687 \times 10^{-5}$ eV/K which is very close to the $k_B = 8.6173 \times 10^{-5}$ eV/K, further validating the classical thermal-activation model [24].

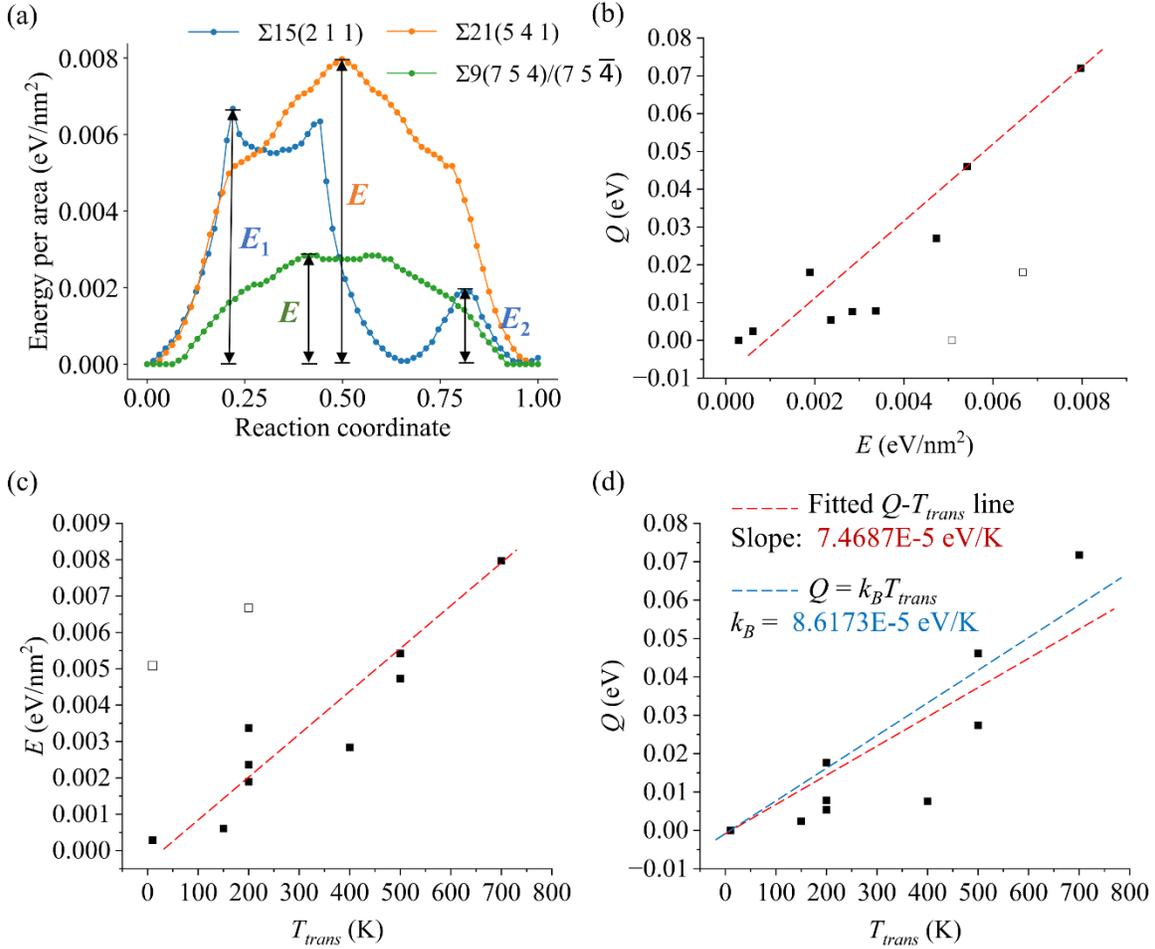

Figure 8 (a) the energy barrier $E$ measured from NEB simulations, and (b) the comparison between $E$ and the apparent activation energy $Q$. (c, d) Plot of (c) $E$ and (d) $Q$ vs. GB thermal behavior transition temperature $T_{trans}$ (□ indicates the second energy barrier $E$ for GBs with two distinct energy barriers)

Indeed, there are limitations to both methods when calculating the activation energy for GB migration. In the classical thermal-activation model, the fitted apparent activation energy $Q$ assumes that the temperature $T$ is much smaller than $T_{trans}$. However, in cases where the GB exhibits anti-thermal behavior, the value of $T_{trans}$ is typically small, making it difficult to satisfy the condition $T \ll T_{trans}$. Additionally, obtaining accurate $M$-$T$ curves at extremely low temperatures can be challenging. Therefore, using the Arrhenius approximation in such cases may introduce potential errors. On the other hand, the energy barrier $E$ calculated using the NEB method has its own limitations. The NEB calculation result strongly depends on the size of the model used, whereas the activation energy for GB migration should be independent of the model size. Moreover, some GBs exhibit a two-step migration process with two energy barriers (Fig. 8a), and apparently one of them is more closely related to the activation energy (Figs. 8b, c). This two-step process has been observed before in literature [50], but the physical mechanisms behind it is still unclear. Nevertheless, there are advantages to using $E$. First, it can be accurately calculated, providing a reliable estimate of the energy barrier. Second, it does not require fitting data across a wide range of temperatures, which can be challenging to obtain. Therefore, $E$ can serve as a useful indicator for predicting the activation energy and thermal behavior of GB migration.

*4.3. The effect of the driving force on the thermal behavior of GB migration*

To investigate the impact of driving force on GB migration, we modified the synthetic driving force code [6] in LAMMPS to be invoked during the energy minimization process. This enabled us to calculate the energy barrier $E$ for GB migration under the influence of synthetic driving forces $\varphi$. As a comparison, we also examined the effect of shear stress $\tau$ on $E$, and an analysis based on the recently proposed concept of GB mobility tensor (see Supplemental Materials for details) revealed that, for causing GB migration at the same mode and at the same velocity, the $\varphi$ and $\tau$ should satisfy the relation: $\tau = \varphi/\beta$. Fig. 9a demonstrates that both the $\varphi$ and $\tau$ can reduce the $E$, and their effects are equivalent. Therefore, for simplicity, our subsequent analysis primarily focused on the $\varphi$.

In Fig. 9b, it is observed that as the driving force increases, the $E$ exhibits a monotonic decrease. This observation is consistent with the decreasing trend of the apparent activation energy $Q$ as the driving force increases, as shown in Fig. 9c. Moreover, the driving force at which the $E$ reaches zero in Fig. 9b coincides with $T_{trans}$ reaching 0 K in Fig. 4, as predicted by Eq. 8. This finding explains the shift of the $T_{trans}$ towards lower temperatures with the increase of the driving force.

The mechanisms discussed above, along with the observations presented in Fig. 5, provide an explanation for the different thermal behaviors exhibited by GBs under varying driving force conditions. When the driving force is low, $T_{trans}$ exhibits a linear relationship with the migration activation energy $Q$. The driving force influences $T_{trans}$ by modifying $Q$, thereby altering the thermal behavior of GB migration. Additionally, a large driving force can activate more migration modes at a lower temperature, which can also change the thermal behavior of GB migration, such as transitioning from anti-thermal to thermally activated behavior (Fig. 4a).

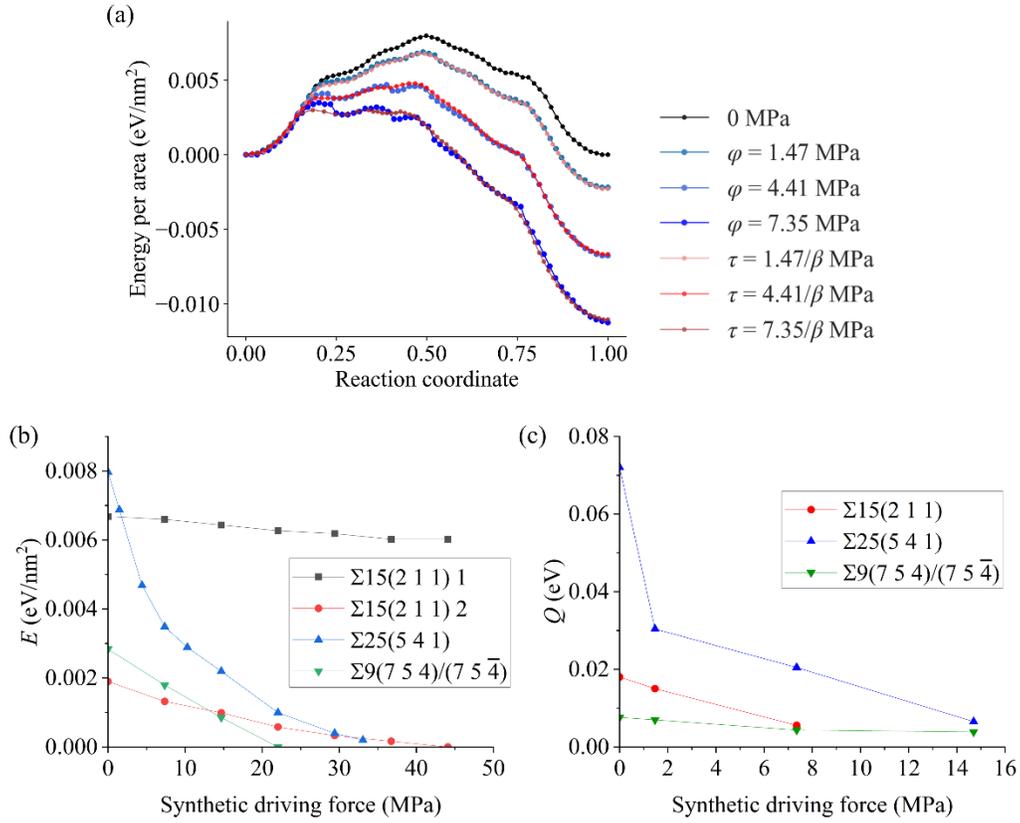

Figure 9 (a) The effect of synthetic driving force $\varphi$ and shear stress $\tau$ on the energy barrier of migration $E$ for the $\Sigma 21(5\ 4\ 1)$ GB. (b) Variation of $E$ with synthetic driving force for $\Sigma 15\ (2\ 1\ 1)$, $\Sigma 9\ (7\ 5\ 4)/(7\ 5\ \bar{4})$, and $\Sigma 21(5\ 4\ 1)$ GBs. (c) Apparent activation energy $Q$ as a function of synthetic driving force for the respective GBs. (Note that the tracking of $Q$ in (c) is limited to a maximum external driving force of 14.7 MPa due to the difficulty in performing Arrhenius fitting at extremely low temperatures.)

Another notable observation is that in Fig. 9b, most of the energy barriers reach zero before the applied synthetic driving force reaches 44.1 MPa. While this value is close to the maximum driving force typically used in experimental situations, it is still relatively small in MD simulations to observe obvious GB migration within a nanosecond timescale. It is worth considering that GBs exhibiting anti-thermal behavior often have small energy barriers for migration, as confirmed by Homer et al. [24]. Therefore, the driving forces used in previous studies [10,11,24] on GB thermal behavior, such as 73.5 MPa and above, may be too large and could have already caused significant changes in the GB migration mechanisms. Furthermore, since the same driving force has different effects on the migration energy barrier in different GBs (Fig. 9b), future studies should consider additional factors. Ideally, the driving force should not exceed the critical value at which the first energy barrier is eliminated if one wishes to explore the GB behavior under typical experimental conditions.

When the driving force reaches extremely high magnitudes, the energy term $\Psi$ introduced by the driving force in Eq. 4 becomes significant and cannot be ignored. By directly referring to Eq. 4, the GB mobility can be expressed as follows:

$$M = \frac{v}{p} = Nbv\,exp\left(\frac{-Q}{k_BT}\right)\left[1 - exp\left(\frac{-\Psi}{k_BT}\right)\right]/P \tag{10}$$

By fitting the data of $\Sigma21(5\,4\,1)$ GB in Fig. 9b, it was found that $M$ decreases exponentially with the driving force $P$, as shown in the insert of Fig. 10. Assuming that the activation energy for GB migration follows the same functional dependence on $P$ as $E$ does, and considering that $\Psi$ is proportional to $P$, all energy terms in Eq. 10 can be expressed in terms of $P$:

$$Q = Q_0 exp(-C_1 P) \tag{11}$$

$$\Psi = C_2 P \tag{12}$$

Here, $Q_0$ is the initial activation energy when the driving force is 0, $C_1$ and $C_2$ are fitted constants. By incorporating these equations into Eq. 10, we can observe from Fig. 10 that the GB mobility $M$ initially increases with P, reaching a peak and then decreasing as $P$ further increases. This behavior aligns with the "anti-driving force" phenomenon observed at low temperatures in Fig. 6. At high temperatures, where $k_BT$ is much larger than both $Q$ and $\Psi$, $M$ becomes a constant and independent of $P$, which is consistent with the findings in Fig. 6. These observations indicate that both the driving force and temperature have a similar effect on GB mobility, and this effect can be described and unified using Eq. 10.

Although the analysis presented so far is based on the synthetic driving force, the comparable effects observed for synthetic driving force and shear stress (Fig. 9a) suggest that these findings should also apply to shear stress conditions.

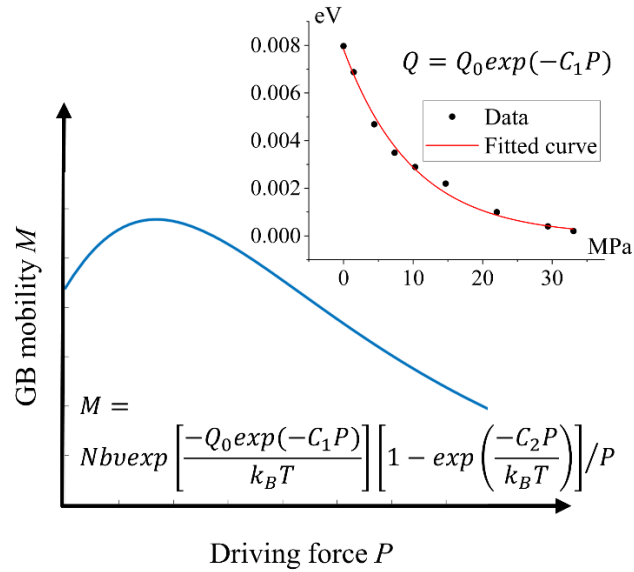

Figure 10 Plot of GB mobility *M* vs. driving force *P* based on Eq. 10

*4.4 Limitations of the current model*

Our study is conducted on an ideal bicrystal system with open ends, which provides a controlled environment to examine the influence of temperature and driving force on GB migration. However, in actual polycrystalline materials, the presence of neighboring grains introduces additional constraints that can impact GB migration behavior. To investigate this, we altered the boundary condition in the x-direction from free surface to periodic and conducted random walk simulations. The results presented in Fig. 11 reveal that changing the boundary condition in the x-direction from free surface to periodic has a significant effect on GB mobility. Below 600K, the GB migration is completely restricted, indicating that the presence of neighboring grains imposes constraints that hinder the GB migration. However, as the temperature increases to 800K, the GB mobilities under both boundary conditions become equal again. This observation aligns with the findings reported by Schratt and Mohles [47].The concept of reconciling grain growth, as investigated by Thomas et al. [51], provides an explanation for this phenomenon. At low temperatures, when only one disconnection mode with strong shear coupling dominates GB migration, the constraints imposed by neighboring grains inhibit the GB migration process. However, at higher temperatures, where multiple migration modes are activated, the constraints can be alleviated by alternatively activating modes with opposite shear coupling signs.

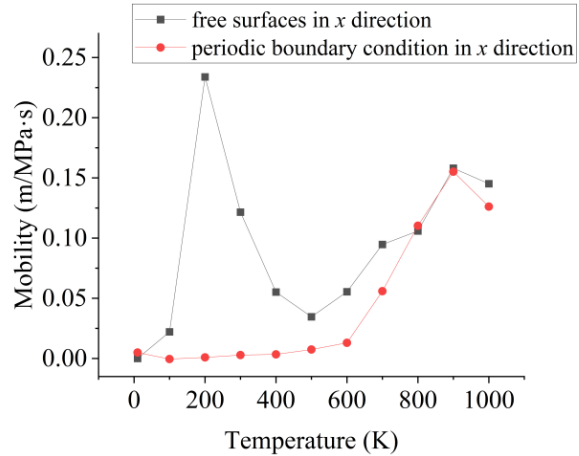

Figure 11 Comparison of the mobility vs. temperature curves for Σ15(2 1 1) GB under different boundary conditions

Indeed, the results presented in Fig. 11 do not contradict the conclusions drawn from the bicrystal model with shrink-wrapped boundary conditions. The constraints imposed by surrounding grains can be considered as an opposite driving force that restrict GB migration by increasing the the activation energy required for GB migration, leading to a significant increase in the $T_{trans}$. Besides, it should be noted that in realistic situations, grains are often much larger than the model used in our study. Consequently, the presence of large grains adjacent to the GB can store a significant amount of elastic energy, mitigating the constraints imposed by neighboring grains. Additionally, real grains possess various mechanisms to alleviate these constraints, including GB rotation [52–54] and the migration of triple junctions[55,56]. Therefore, despite the simplifications inherent in the bicrystal system, studying it remains informative and provides valuable reference for understanding GB behavior in real-world cases.

Another limitation of the classical thermal-activation model is its assumption that only one migration mode dominates GB migration. However, in polycrystalline materials, the phenomenon of reconciling GB migration necessitates the consideration of multiple migration modes [51]. Disconnection theory [30] proposes that multiple modes can be simultaneously activated at different locations along the GB in the form of disconnections, which are line defects combining step height and dislocation characteristics. This assumption has been supported by experimental observations [57]. Chen et al. [7] hypothesize that the GB mobility is the sum of the mobilities of each mode, expressed as:

$$M = \sum_m M_m \tag{13}$$

Here, $M_m$ represents the mobility equation (Eq. 5) for the $m$-th migration mode. However, accurately measuring the constant terms and activation energies in Eq. 5 for all the modes remains a challenge.

## 5. Conclusion

By conducting systematic atomistic simulations of GB migration using both dynamic and static approaches, we have derived several key conclusions:

- The anti-thermal behavior of GB mobility is an intrinsic nature of GBs, and the transition temperature ($T_{trans}$) in thermal behavior, at which the mobility of GB is at its maximum, exhibits a linear relationship with the activation energy for GB migration. This finding provides support for both Homer's classical thermal-activation model [24,25] (Eq. 5) and Chen et al.'s disconnection-based theory [7] (Eq. 7).
- The synthetic driving force can reduce the activation energy required for GB migration, leading to a shift in the transition temperature ($T_{trans}$) to lower values. Moreover, a higher driving force can activate more migration modes at lower temperatures, thereby influencing the thermal behavior of the GB. We suggest an optimal range of driving forces that should be considered in future atomistic simulations investigating the GB thermal behavior under typical experimental conditions.
- Through our quantitative analysis of the effects of the synthetic driving force and shear stress on the energy barrier for GB migration, we have observed that these two driving forces are equivalent in their ability to lower the energy barrier. This means that the conclusions drawn from studying the synthetic driving force can also be applied to the case of shear stress.

- We have refined the thermal-activation model by incorporating the influence of the driving force on the activation energy. This modified model accurately captures the observed "anti-driving force" trend in GB mobility resulting from changes in the driving force.

These findings contribute to our understanding of GB migration and provide insights into the interplay between temperature, driving force, and GB mobility, which is essential for GB engineering and modern industry.

**Acknowledgment**

The authors thank Dr. David L Olmsted for sharing the 388 Ni GB structure database, Dr. Penghui Cao for sharing the SLME simulation code, and Ali Kerrache for providing technique support at Compute/Calcul Canada. ChatGPT was used to polish the final writing of the paper under supervision. This research was supported by NSERC Discovery Grant (RGPIN-2019-05834), Canada, and the use of computing resources provided by WestGrid and Compute/Calcul Canada.

# Supplementary materials

# for

# Driving force induced transition in thermal behavior of grain boundary migration in Ni

Xinyuan Song and Chuang Deng*

Department of Mechanical Engineering, University of Manitoba, Winnipeg, MB R3T 2N2, Canada

* Corresponding author: Chuang.Deng@umanitoba.ca

## S1. Dichromatic pattern analysis for Σ15 (2 1 1), Σ21 (5 4 1), and Σ9 (7 5 4)/(7 5 $\bar{4}$) GBs

According to the unified grain boundary (GB) kinetics model proposed by Han et al. [1], GB migration is mediated by the nucleation and movement of disconnections, which are line defects characterized by both height $h$ and Burgers vector $\boldsymbol{b}$. The activation energy for disconnection nucleation increases with higher values of $\boldsymbol{b}$ or $h$. The dichromatic pattern of GB is a visualization technique that can be obtained by extending the grains on either side of the GB throughout the entire space. It allows for the analysis of different disconnection modes. To perform the analysis, one can displace the atoms on one side of the GB on the dichromatic pattern by a vector of $\boldsymbol{b}$ to restore the coincidence-site lattice (CSL) pattern. This displacement reveals the possible disconnection modes and corresponds to the migration of the GB over a distance of $h$, as illustrated in Fig. S1. It is important to note that the dichromatic pattern analysis is limited to the GB plane that exhibits the CSL pattern, specifically, the *x-z* plane for for Σ15 (2 1 1) and Σ21 (5 4 1) GBs and *x-y* plane for for Σ9 (7 5 4)/(7 5 $\bar{4}$) GB, aligning with the observations made in our study that shear coupling occurs exclusively in the plane exhibiting the CSL pattern.

The dichromatic pattern of the Σ15 (2 1 1) GB clearly shows the disconnection mode with the lowest activation energy, and its shear coupling factor $\beta = \boldsymbol{b}/h = 0.89$ aligns with the simulation results (Fig. 3a in the main text). As the temperature and driving force increase, the mode with opposite values of $\boldsymbol{b}$ can be activated, leading to a decrease in $\beta$.

In the case of the Σ21 (5 4 1) GB, the mode highlighted in red exhibits the same $\beta$, i.e. 0.39, as the one computed from the simulation results (Fig. 3c in the main text). This suggests that the value of the $\boldsymbol{b}$ appears to be predominantly associated with the activation energy of the disconnection mode in this particular GB system.

For Σ9 (7 5 4)/(7 5 $\bar{4}$) GB, the dichromatic pattern reveals two distinct disconnection modes. The first type involves modes with a $\boldsymbol{b}$ perpendicular to the GB plane, resulting in pure GB migration without shear coupling. The second type consists of pairs of modes with similar $h$ but opposite $\boldsymbol{b}$. These modes also indicate GB migration without shear coupling. These observations align with the results shown in Fig. 3b in the main text.

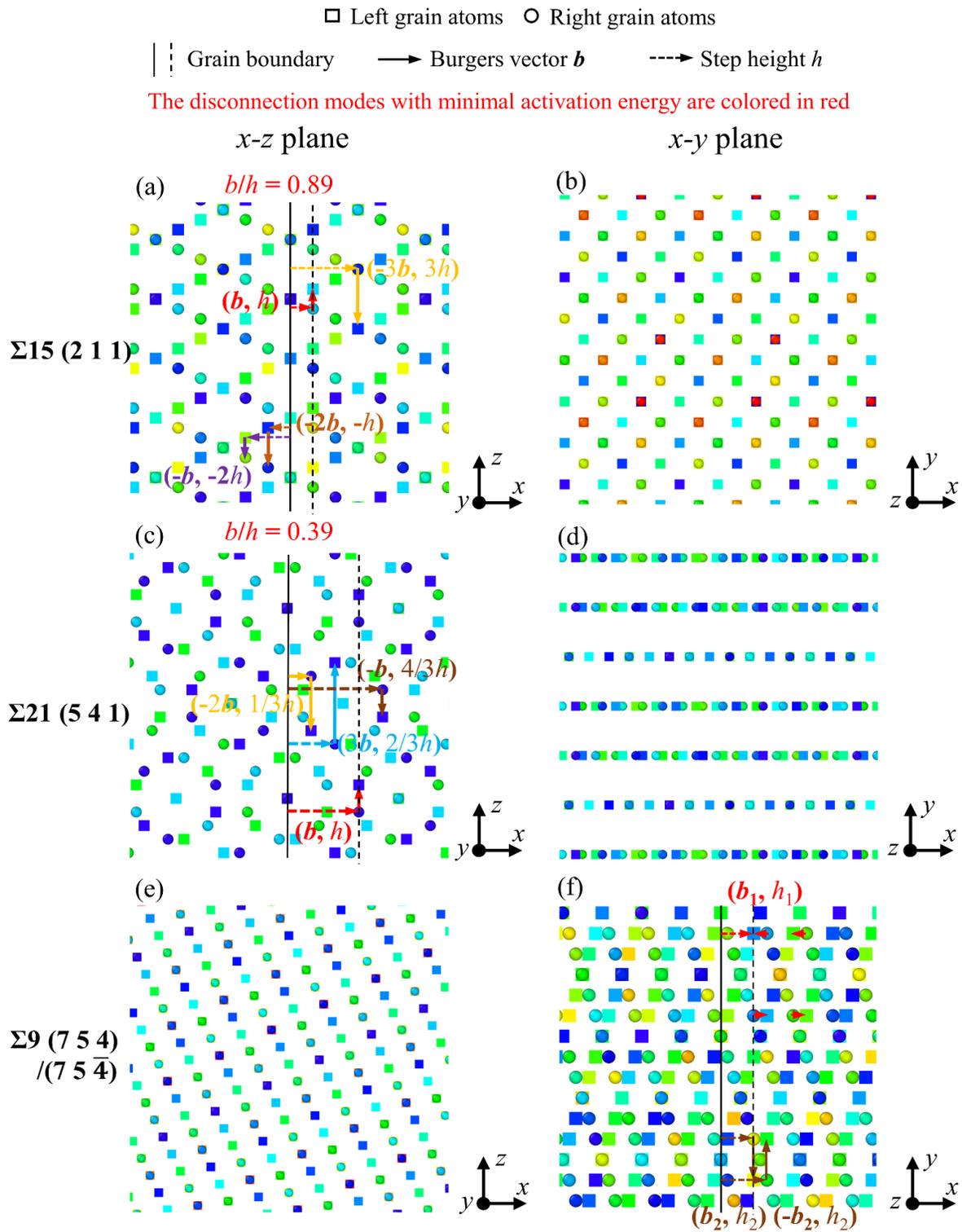

Figure S1 Dichromatic pattern analysis for (a, b) Σ15 (2 1 1), (c, d) Σ21 (5 4 1), and (e, f) Σ9 (7 5 4)/(7 5 $\bar{4}$) GBs. The possible disconnection mode associated with the lowest activation energy is marked in red.

The solid line indicates the position of the GB (the plane with the least atoms and lowest energy), and the dashed line indicates the GB position after the lowest energy mode has been activated.

## S2. Plots of GB mobility vs. temperature curves of GBs in Table 1 determined by random walk method

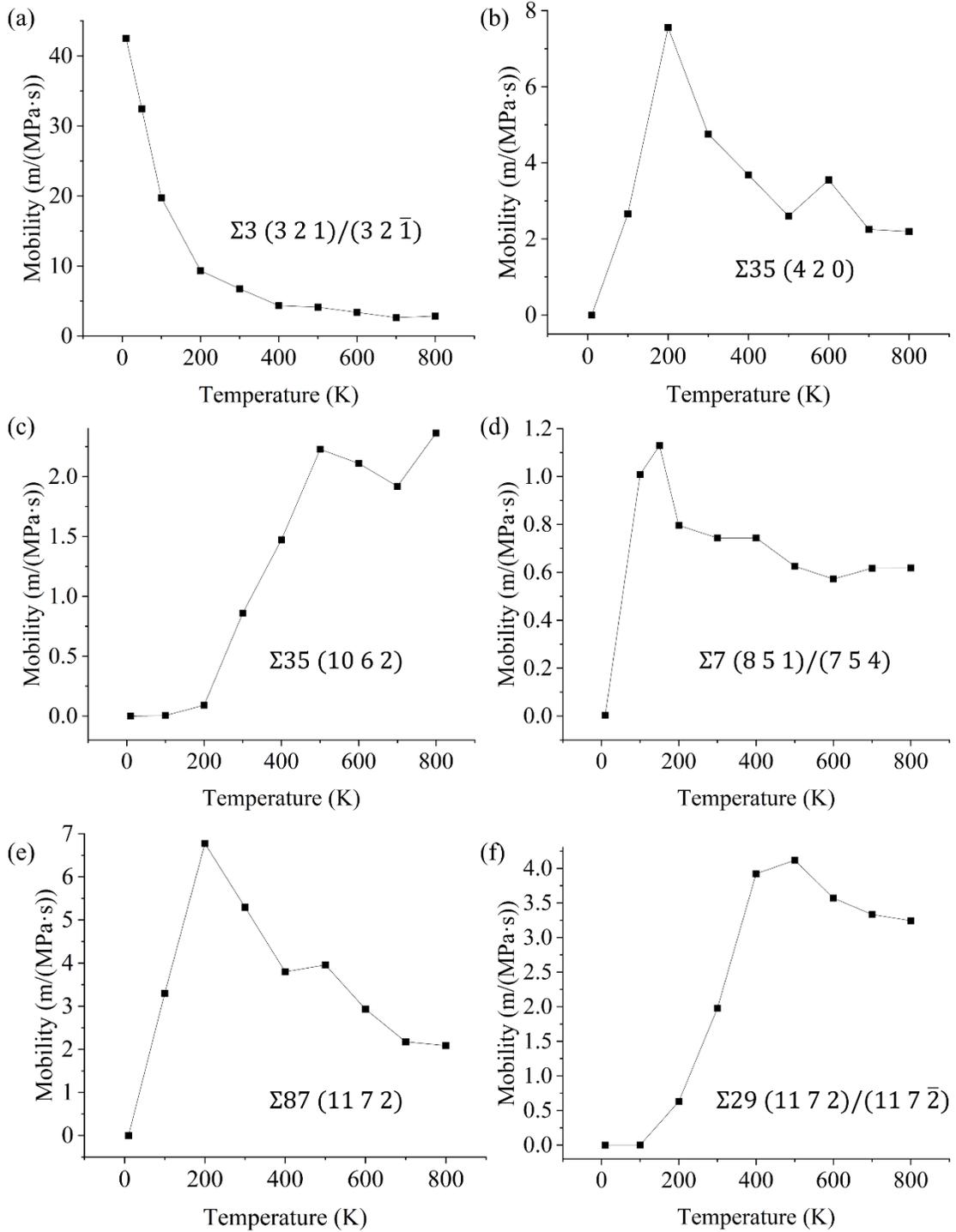

Figure S2 Plots of GB mobility vs. temperature curves of GBs in Table 1 determined by random walk method

## S3. Calculation of energy barrier for GB migration

The energy barrier for GB migration is determined using the a combined approach of self-learning metabasin escape (SLME) [2–4] and nudged elastic band (NEB) [5–7] methods. Initially, the SLME method is employed to search for neighboring local minima in the potential energy surface (PES) of the current atomic configuration. In this process, a penalty function ϕ is introduced to the PES, which is defined by:

$$\phi = h \exp\left[\frac{-(r-s)^2}{2w^2}\right] \tag{1}$$

where $r$ is the coordination of the atomic configuration on 3D PES, $s$ is the current atomic configuration, $h$ is the height of the penalty function, and $w$ is the half-width of the penalty function. The SLME approach can self-learn the current penalty function $\phi$ on the fly, and generate a new penalty function and add it to the PES until the system jumps to a neighboring local minimum, resulting in a small migration of GB, as illustrated in Fig. S3.

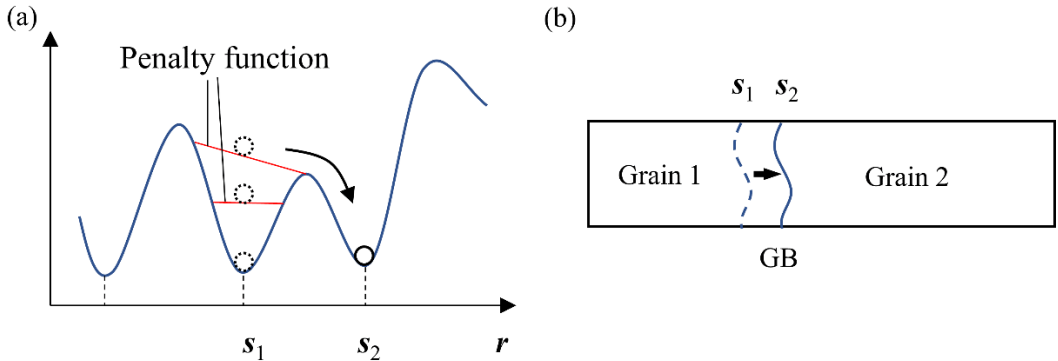

Figure S3 Illustration of the principle of SLME approach

However, the SLME approach is not highly accurate in determining the saddle point. Therefore, after obtaining the atomic configurations of two adjacent local minima, we employed the nudged elastic band (NEB) method with a parallel nudging force of 1.0 eV/Å to calculate the saddle point between the two adjacent local minima. The height of the saddle point, divided by the GB area, was considered as the energy barrier for GB migration. The division of the GB area helps to eliminate the size effect in the NEB method and provides a more meaningful measure of the energy barrier for GB migration.

## S4. GB mobility under varying synthetic driving forces (58.8 to 440.97 MPa) and temperatures

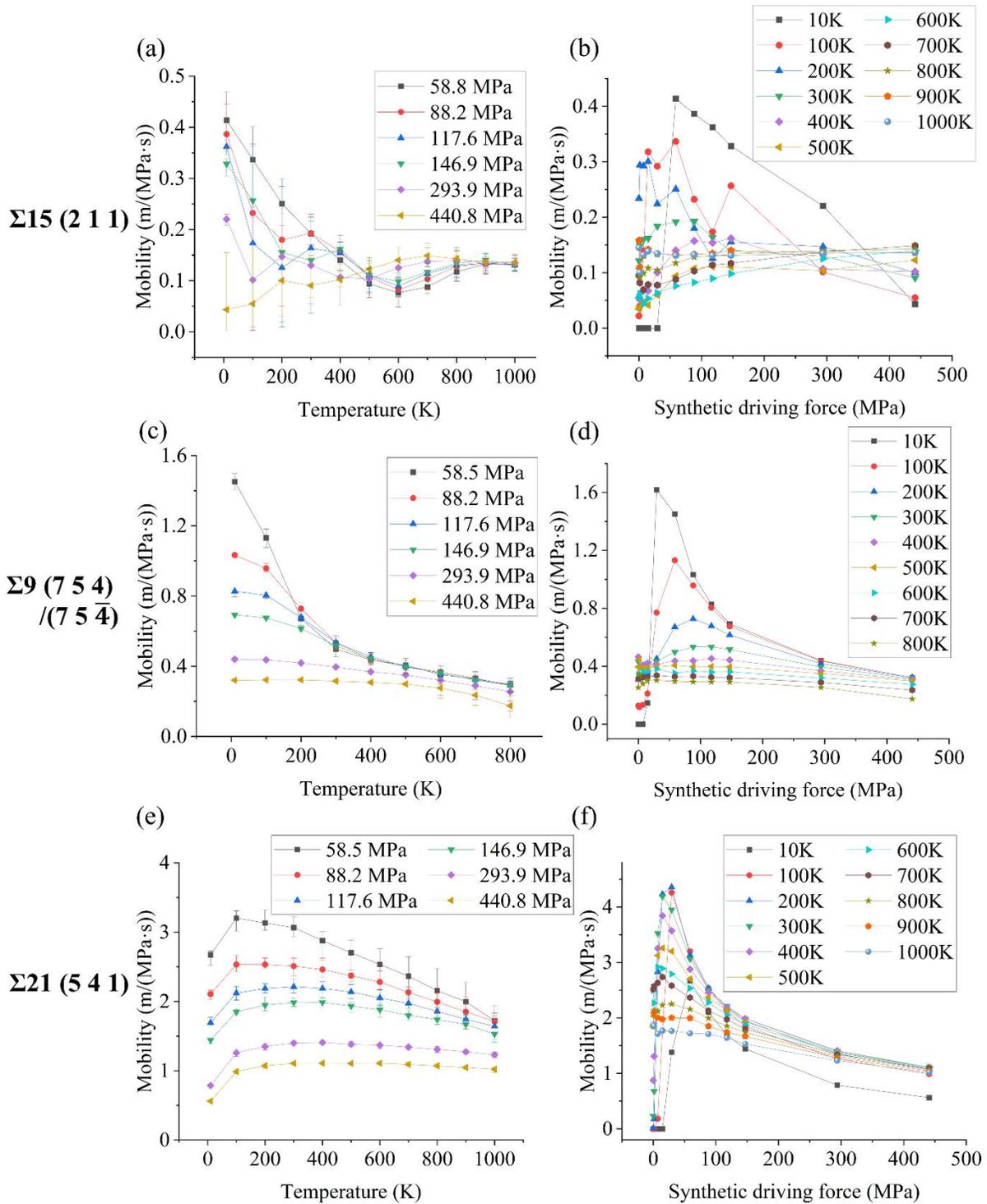

Figure S4 Effect of synthetic driving forces and temperature on GB mobility. (a, c, e) GB mobility vs. temperature curves under various synthetic driving forces ranging from 58.8 to 440.97 MPa. (b, d, f) GB mobility vs. synthetic driving force curves at different temperatures.

## S5. Equivalence between shear stress and synthetic driving force

According to the recently proposed concept of the GB mobility tensor [8], for a GB with shear coupling specifically in the *x-z* plane, the kinetic equation for GB migration can be expressed as:

$$\begin{bmatrix} v_x \\ v_z \end{bmatrix} = \begin{bmatrix} M_{xx} & M_{xz} \\ M_{zx} & M_{zz} \end{bmatrix} \begin{bmatrix} \varphi \\ \tau \end{bmatrix} \qquad (2)$$

This equation represents the relationship between the components of the GB velocity and different types of driving forces ($\varphi$, $\tau$), where $M_{xx}$, $M_{xz}$, $M_{zx}$, and $M_{zz}$ are the corresponding mobilities.

When there is only one migration mode activated, the elements in the mobility tensor should satisfy the relation: $M_{zx} = M_{xx} \cdot \beta$ and $M_{xz} = M_{zz}/\beta$, where $\beta$ can be analyzed from dichromatic pattern, as shown in Fig. S1. According to the Onsager reciprocal relations [8–10], the GB mobility tensor should be symmetric, implying $M_{xz} = M_{zx} = M_{xx} \cdot \beta$. Therefore, when the synthetic driving forces $\varphi$ and shear stress $\tau$ lead to the same GB velocity in *x* direction, $\varphi$ and $\tau$ should follow the relation $\tau = \varphi/\beta$.